# Influences of stoichiometry deviations on geometry and electronic structures of TiO$_2$ anatase (101)

Qinggao Wang[1, 3] and Artem R. Oganov[2, 1, 4, 5]

1 Moscow Institute of Physics and Technology, 9 Institutskiy Lane, Dolgoprudny City, Moscow Region, 141700, Russia

2 Skolkovo Institute of Science and Technology, Skolkovo Innovation Center, 5 Nobel St., Moscow 143026, Russia

3 Department of Physics and Electrical Engineering, Anyang Normal University, Anyang, Henan Province, 455000, the People's Republic of China

4 Department of Geosciences and Center for Materials by Design, Stony Brook University, Stony Brook, New York 11794, USA

5 School of Materials Science and Engineering, Northwestern Polytechnical University, Xi'an, Shanxi 710072, People's Republic of China

**Abstract** We systematically investigated native defects and reconstructions of TiO$_2$ anatase (101) using USPEX and VASP codes. For lightly reduced samples, the coverage of O vacancies at surface depends on its concentration in bulk, while Ti interstitials prefer to dissolve in bulk. Atomic configurations of reconstructed TiO$_2$ anatase (101) structures were given for severe reduced samples at the first time. Increasing the reduction degree, the Fermi level can be modulated in the whole bandgap. For O-rich samples, due to the indirect-to-direct bandgap transition and bandgap narrowing, O interstitials are responsible for visible light photoactivity. In a word, stoichiometry deviations of TiO$_2$ anatase (101) have broad implications for photocatalytic reactions.

TiO$_2$ is one of the versatile semiconductors, whose fascinating properties closely relate to bulk and surface structure [1]. Generally, it has been agreed that anatase polymorph has the highest photoactivity followed by rutile and brookite polymorph [2]. {101} facets are abundant [3, 4], however, structures of (101) surface are still controversial.

Experimentally, scanning tunneling microscopy (STM) and low energy electron diffraction (LEED) investigations revealed the (1×1) periodicity of TiO$_2$ anatase (101) surface [5, 6], and such a periodicity usually corresponds to a bulk-truncated structure. In contrast, x-ray photoelectron spectroscopy indicated that a Ti-rich $TiO_2$ anatase (101) structure also has the (1×1) periodicity [7]. Accordingly, the (1×1) periodicity also corresponds to reconstructed structures at O-deficient conditions, but atomic configurations have not been resolved.

Native defects also correspond to stoichiometry deviations, such as O vacancies. By comparison, defected rutile $TiO_2$ was more extensively investigated. Lee *et al.* reported that both O vacancies and Ti interstitials are relevant for O-deficiency, but O vacancies are dominated defects at lightly reduced conditions [8]. Ti interstitials [9] and O vacancies [9, 10] of $TiO_2$ anatase (101) have been investigated, but the relation between their surface coverage and concentration in bulk has not been discussed.

O interstitials strongly bind to lattice O atoms of anatase $TiO_2$ with forming $(O_2)_O$ species [11]. Experimentally, Etacheri *et al.* reported that excess O atoms enhance visible light photoactivity attributed to a bond gap narrowing [12]. Accordingly, $(O_2)_O$ species cause visible light photoactivity. Furthermore, Setvin *et al.* reported that $(O_2)_O$ species form at $TiO_2$ anatase (101) surface through an reaction of $O_2$ molecules with subsurface O vacancies [13]. However, electronic structure of O interstitials has not been theoretically investigated.

In this paper, several (1×1) reconstructions were discovered at the first time. And native defects were systematically investigated. Finally, we calculated band structures of meticulously selected structures.

Possible reconstructions of $TiO_2$ anatase (101) were explored using the USPEX package [14-17]. Four multiplications of unit cell were allowed, and maximally four Ti and eight O atoms were allowed for per unit cell. Each of supercells consisted of a slab having four $TiO_2$ layers and a vacuum layer of 15 Å. Only the topmost region maximally deep into 3.5 Å was allowed to relax using the VASP code [18-20]. Low-energy structures were selected for post-processing. Therein, slab models were increased to five or six $TiO_2$ layers, and only the bottom ones were fixed.

Native defects at surface or in bulk were systematically investigated by performing spin-polarized VASP calculations. The (2×2) slab models, consisted of four TiO$_2$ layers and a vacuum layer of 15 Å, were adopted. For bulk, both (2×2×1) and (4×2×1) supercells, consisted of 16 and 32 TiO$_2$ unit cells, respectively, were adopted.

The electron exchange and correlation were treated within the generalized gradient approximation (GGA) with the addition of a Hubbard U term of 2.5 eV. This value accords with delocalized electrons of anatase TiO$_2$ [21]. And thus, the reported errors in energetics [9], could be avoided. Orbitals of valence electrons were expanded in plane waves, and the projector augmented wave method was applied to treat core electrons [22]. Γ-centered 2π×0.09 Å$^{-1}$ sampling was adopted for the Brillouin zone integration. Structural relaxations progressed until the forces on each of atoms are less than 0.001 eV/Å. Moreover, dipole corrections were adopted to cancel interactions between slabs and periodic boundary [23, 24].

Formation energy reflects the stability of native defects, which is defined as

$$E_f = E_{TiO_2-D} - n_{Ti}E_{TiO_2} - \mu_O(n_O - 2n_{Ti}), \tag{1}$$

where $E_{TiO_2-D}$ is the total energy of a defected structure. $E_{TiO_2}$ is the total energy of bulk anatase TiO$_2$ averaged to per unit cell. $n_O$ and $n_{Ti}$ are the numbers of O and Ti atoms, respectively. $\mu_O$ is the chemical potential of O atoms.

The equilibrium growth of anatase TiO$_2$ depends on chemical potential, which satisfy following relations: (a) $\mu_O \leq \frac{1}{2}\mu_{O_2}$, (b) $\mu_{Ti} \leq \mu_{Ti}^{bulk}$, (c) $\mu_{Ti} + 2\mu_O = \mu_{TiO_2}^{anatase}$, and (d) $2\mu_{Ti} + 3\mu_O = \mu_{Ti_2O_3}^{bulk}$. Accordingly, $\mu_O$ must fall into the region of $-8.93eV \leq \mu_O \leq -4.69eV$, as illustrated in supporting material.

**Reconstructed TiO$_2$ anatase (101) structures** In total, 1777 structures were explored during USPEX calculations. Low-energy (1×1) reconstructed TiO$_2$ anatase (101) structures were discovered at the first time. Their stability is investigated through recasting the formula (1),

$$E_f = E_{TiO_2(101)} - n_{Ti}E_{TiO_2} - \mu_O(n_O - 2n_{Ti}). \qquad (2)$$

Accordingly, the unreconstructed surface and "added-TiO+$V_O$" reconstruction can become stable (Fig.1). And, metastable "added-TiO" and "added-TiO+2$V_O$" reconstructions are predicted. In the light of chemical potential, the added-TiO+$V_O$ reconstruction can become stable only for extreme reduced samples.

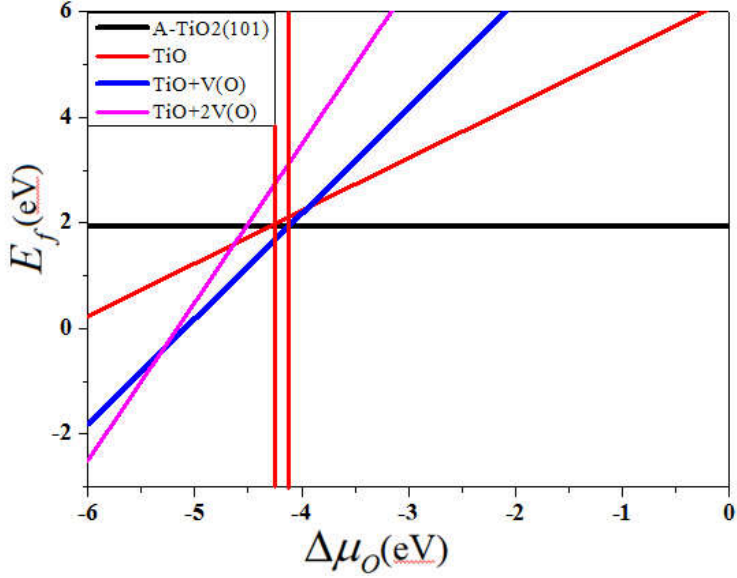

Fig.1 (color online) Stability of TiO$_2$ anatase (101) structures changing with $\mu_O$.

Reconstructed structures have the (1×1) periodicity, matched with experimental observations [5, 7, 25, 26]. This finding surpasses a common understanding on surface reconstruction, i.e., the (1×1) surface was regarded as the bulk-terminated structure [1, 5]. However, the bulk-terminated structure is stable nearly at the whole region of allowed chemical potential. And thus, kinetically accessible defects are important to understand functions of anatase TiO$_2$[27], although they are metastable in thermodynamics.

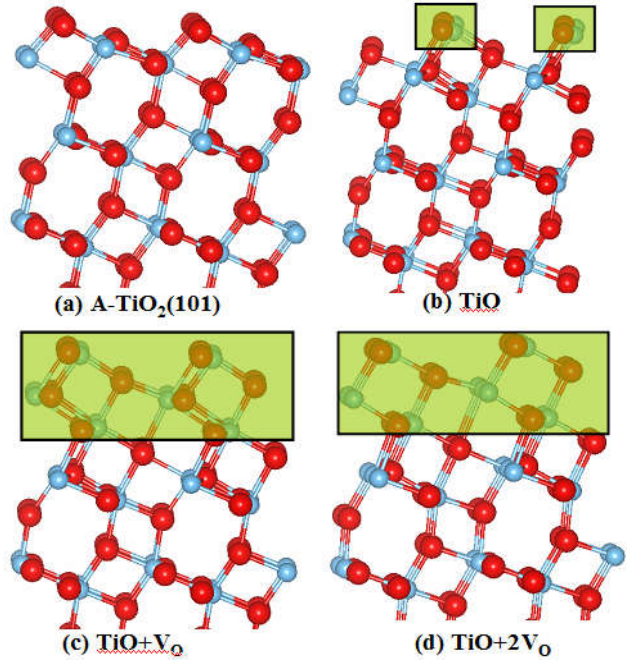

Fig.2 (color online) Side views of (a) the unreconstructed surface, and predicted (b) added-TiO, (c) added-TiO+$V_O$, (d) added-TiO+2$V_O$ reconstructions. Their features are highlighted.

In Fig.2a, the unreconstructed surface shows two- and three-fold O atoms, as well as five- and six-fold Ti atoms, as reported in literatures [25, 28]. Different from unreconstructed surface, the added-TiO reconstruction owns TiO stripes along the [010] direction (Fig.2b). The added-TiO+$O_v$ reconstruction is further introduced an O vacancy with forming the Frenkel pair (i.e., the $V_{Ti}$-$Ti_i$ pair) along the [010] direction (Fig.2c). Experimentally, He *et al.* found that subsurface O vacancies tend to form linear arrays best explained by Frenkel pairs [7]. By comparison, we give an updated result, i.e., both Frenkel pairs and added-TiO stripes are included. Furthermore, the added-TiO+2$V_O$ reconstruction (Fig.2d) is more severe reduced.

**Native defects in bulk** Atomic configurations of defected anatase $TiO_2$ were built according to published results [9, 11]. At the extreme O-rich condition, the $E_f$ value of per Ti interstitial is 8.86 eV adopting the (2×2×1) supercell. But at the extreme Ti-rich condition, the $E_f$ value becomes 0.38 eV. Accordingly, it is easy to form Ti interstitials with decreasing O chemical potential. This is the same as that of O vacancies (Table 1). In contrast, O interstitials prefer to form at the extreme O-rich condition, since the $E_f$ value is the smallest (Table 1). To our knowledge, anatase TiO$_2$ is quite different from rutile polymorph, and no one reported O interstitials in rutile TiO$_2$.

Table 1 Formation energies of per Ti interstitial, O vacancy and interstitial. Both (2×2×1) and (4×2×1) supercells were adopted.

| Defects\ | $E_f$ (2×2×1) | | $E_f$ (4×2×1) | |
|---|---|---|---|---|
| | O-rich | Ti-rich | O-rich | Ti-rich |
| Ti$_{int}$ | 8.86 | 0.38 | 8.44 | -0.04 |
| O$_{vac}$ | 4.71 | 0.47 | 4.15 | -0.09 |
| O$_{int}$ | 0.84 | 5.08 | 0.09 | 4.33 |

The formation of native defects is easier at a low concentration. For example, the $E_f$ value of per O interstitial is 0.84 eV at the extreme O-rich condition adopting the (2×2×1) supercell, while it is 0.09 eV adopting the (4×2×1) supercell.

Under O-poor conditions, Lee *et al.* reported that the $E_f$ value of per Ti interstitial in rutile TiO$_2$ (0.3 eV) is smaller than that of per O vacancy (1.9 eV) [8], and thus Ti interstitials are dominated defect. In contrast, O interstitials are dominated defects of anatase TiO$_2$ at O-rich conditions, since its $E_f$ value (0.09 eV) is smaller compared with those of per O vacancy and Ti interstitial (Table 1). With decreasing O chemical potential, O vacancies dominate O-deficiency for lightly reduced samples (Table 1). At extreme Ti-rich conditions, O vacancies and Ti interstitials codetermine O-deficiency, and their $E_f$ values are comparable (Table 1).

**Native defects at surface** The coverage of native defects at surface should be in equilibrium with their concentration in bulk. Atomic configurations of defected TiO$_2$ anatase (101) were consciously built according to published results [9, 29]. Only surface and the most stable subsurface defects were considered.

At the extreme O-rich condition, the $E_f$ value is 4.69 (4.64) eV for per O vacancy at surface (subsurface). In agreement with previous results [7, 9, 10, 29], the formation of subsurface O vacancies is a little more preferable. At the same chemical potential, the $E_f$ value of surface O vacancies is in between those in bulk. For example, at the extreme O-rich condition, the $E_f$ value of per surface O vacancy (4.69 or 4.64 eV) is smaller than that in bulk adopting the (2×2×1) supercell (4.71 eV), but larger than that with adopting the (4×2×1) supercell (4.15 eV). Consequently, the coverage of O vacancies at surface closely relates to its concentration in bulk.

Table 2 Formation energies of per O vacancy, Ti interstitial and O interstitial. Both surface and the most preferable subsurface sites of TiO$_2$ anatase (101) were considered.

| Defects\ | $E_f$ (surface) | | $E_f$ (subsurface) | |
|---|---|---|---|---|
| | O-rich | Ti-rich | O-rich | Ti-rich |
| O$_{vac}$ | 4.69 | 0.46 | 4.64 | 0.41 |
| Ti$_{int}$ | 9.27 | 0.79 | 9.09 | 0.61 |
| O$_{int}$ | 0.61 | 4.85 | 1.26 | 5.50 |

For the TiO$_2$ anatase (101) surface, Ti interstitials are quite different from O vacancies. The $E_f$ value of per subsurface Ti interstitial (9.27 eV) is smaller than that of surface Ti interstitials (9.09 eV), agreed with a previous result [9]. At the same chemical potential, the $E_f$ value of per Ti interstitial at surface (9.27 or 9.09 eV) is much larger than that in bulk (Table 1). Accordingly, Ti interstitials prefer to dissolve in bulk rather than localizing at surface. So, O vacancies are dominated defects of lightly reduced TiO$_2$ anatase (101).

According to previous results [11, 12], O interstitials enhance visible light photoactivity due to bond gap narrowing. At the extreme O-rich condition, the $E_f$ value is 0.61 (1.26) eV for per surface (subsurface) O interstitial. Different form O vacancies, O interstitials prefer to locate at surface. Furthermore, the $E_f$ value of per O interstitial at surface (0.61 eV) is in between those values in bulk (0.84 and 0.09 eV). Consequently, the coverage of O interstitials at surface depends on its concentration in bulk.

For lightly reduced samples, the coverages of O vacancies and interstitials at surface correlate with their concentrations in bulk, respectively, while Ti interstitials prefer to dissolve in bulk.

**Band structure** Since band structures closely relate to photocatalytic reactions, we investigate them for defected and reconstructed $TiO_2$ anatase (101) structures.

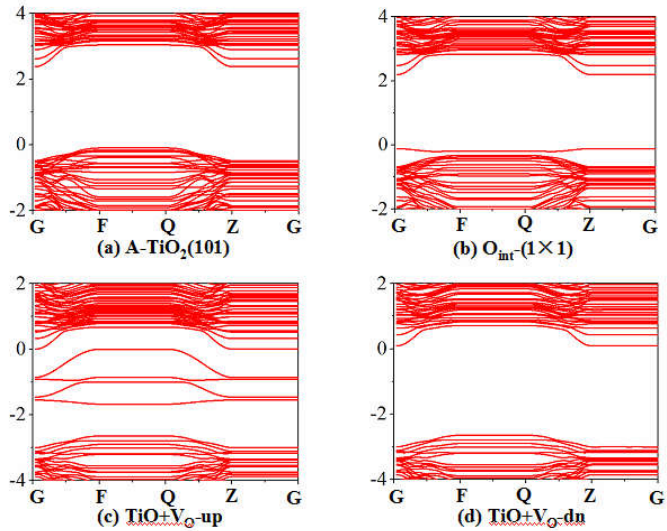

Fig.3 (color online) Band structures of (a) the unreconstructed surface and (b) the $TiO_2$ anatase (101) owning an O interstitial. (c) and (d) are spin-up and -down band structures of the $TiO+O_v$ reconstruction, respectively. In these figures, the Fermi level was moved to 0 eV.

Compared with band structures of A-$TiO_2$(101) and $O_{int}$-(1×1), new valence states at edge appear due to O interstitials. Clearly, the unreconstructed surface has an indirect band gap, and thus the optical absorption near edge must involve phonons. In contrast, the $O_{int}$-(1×1) has a direct band gap, so its light absorption is purely an electron-photon interaction. Accordingly, O interstitials, causing valence states at edge, lead to the indirect-to-direct bandgap transition. Furthermore, new valence states are also resulted in a bandgap narrowing, and more O interstitials should correspond to a narrower bandgap. Consequently, O-rich anatase $TiO_2$ has visible light photoactivity. By far, we have rationalized an experimental result that O–rich anatase $TiO_2$ has visible light photoactivity [12]. However, no O-rich rutile $TiO_2$(110) sample was reported.

In Fig.3c, the Ti-rich $TiO+O_V$ reconstruction has gap states, corresponding to electrons localized at $Ti^{3+}$ ions. Interestingly, the Fermi level of TiO+$O_V$ reconstruction, corresponding to the chemical potential of electrons, was moved to the edge of conduction band. However, the Fermi level of the most severe reduced rutile $TiO_2$(110) is in bandgap [17]. Compared with the rutile $TiO_2$(110) surface, the Fermi level of $TiO_2$ anatase (101) is easier to be modulated, and thus its charge exchange with ambient environment is easier. This deduction is in agreement with an experimental result that electrons of anatase $TiO_2$ are more delocalized than rutile polymorph [21].

In summary, stoichiometry deviations of $TiO_2$ anatase (101) surface are significant for photocatalytic reactions. Successively, O interstitials and vacancies are dominated defects with decreasing O chemical potential, respectively. Due to an indirect-to-direct bandgap transition and bandgap narrowing, O interstitials are responsible for visible light photoactivity. For severe reduced samples, (1×1) reconstructed structures were predicted at the first time, although the unreconstructed surface is stable nearly at the whole region of chemical potential. Interestingly, the Fermi level, corresponding to chemical potential of electrons, can be modulated at the whole bandgap through changing reduction degree. Compared with rutile $TiO_2$, electrons of defected and reconstructed anatase $TiO_2$ structures are more delocalized [21], and to our knowledge, O-rich rutile $TiO_2$ was not reported. In a word, our results will play an important role for understanding fascinating functions of anatase $TiO_2$.

**Acknowledgements** Qinggao Wang thanks Jianhua Zhao, Huafeng Dong and Congwei Xie for valuable discussions. This research is supported by the Government of the Russian Federation (No. 14.A12.31.0003), DARPA (Grant No. W31P4Q1210008), the National Natural Science Foundation of China (Grant No. 11504004) and the "5top100" postdoctoral fellowship conducted at Moscow Institute of Physics and Technology.